# The Large Array Survey Telescope—Science Goals

S. Ben-Ami[1], E. O. Ofek[1], D. Polishook[2], A. Franckowiak[3], N. Hallakoun[1], E. Segre[2], Y. Shvartzvald[1], N. L. Strotjohann[1], O. Yaron[1], O. Aharonson[4], I. Arcavi[5], D. Berge[6], V. Fallah Ramazani[3], A. Gal-Yam[1], S. Garrappa[3], O. Hershko[2], G. Nir[7], S. Ohm[6], K. Rybicki[1], I. Sadeh[6], N. Segev[1], Y. M. Shani[1], Y. Sofer-Rimalt[1], and S. Weimann[3]

[1] Department of Particle Physics and Astrophysics, Weizmann Institute of Science, 76100 Rehovot, Israel
[2] Department of Physics Core Facilities, Weizmann Institute of Science, 76100 Rehovot, Israel
[3] Faculty of Physics and Astronomy, Astronomical Institute (AIRUB), Ruhr University Bochum, D-44780 Bochum, Germany
[4] Department of Earth and Planetary Sciences, Weizmann Institute of Science, 76100 Rehovot, Israel
[5] The School of Physics and Astronomy, Tel Aviv University, Tel Aviv 69978, Israel
[6] Deutsches Elektronen-Synchrotron (DESY), D-15738 Zeuthen, Germany
[7] University of California, Berkeley, Department of Astronomy, Berkeley, CA 94720, USA

## Abstract

The Large Array Survey Telescope (LAST) is designed to survey the variable and transient sky at high temporal cadence. The array is comprised of 48 F/2.2 telescopes of 27.9 cm aperture, coupled to full-frame backside-illuminated cooled CMOS detectors with 3.76 $\mu$m pixels, resulting in a pixel scale of $1''\!\!.25$. A single telescope with a field of view of 7.4 deg$^2$ reaches a 5$\sigma$ limiting magnitude of 19.6 in 20 s. LAST 48 telescopes are mounted on 12 independent mounts—a modular design which allows us to conduct optimized parallel surveys. Here we provide a detailed overview of the LAST survey strategy and its key scientific goals. These include the search for gravitational-wave (GW) electromagnetic counterparts with a system that can cover the uncertainty regions of the next-generation GW detectors in a single exposure, the study of planetary systems around white dwarfs, and the search for near-Earth objects. LAST is currently being commissioned, with full scientific operations expected in mid 2023. This paper is accompanied by two complementary publications in this issue, giving an overview of the system and of the dedicated data reduction pipeline.

*Unified Astronomy Thesaurus concepts:* Telescopes (1689); Automated telescopes (121); Optical telescopes (1174); Sky surveys (1464); Transits (1711); Transient detection (1957); Supernovae (1668); Gravitational waves (678)

## 1. Introduction

Systematic mapping of the sky is at the core of modern astronomy. Sky surveys are delivering more scientific data than ever before, allowing us to study the Universe and its constituents, discover new types of objects and phenomena, and conduct unprecedented tests of the basic laws of physics. Today's large-scale digital sky surveys are transforming the way astronomy is done, allowing almost real-time access to state-of-the-art observations for an unprecedented number of scientists around the globe.

Benefiting from technological advancements (e.g., large format CCDs), numerous wide-field imaging survey telescopes have been constructed in the past two decades, and new machines are being built. Examples include the Palomar Transient Factory (PTF; Law et al. 2009), the Pan-STARRs survey (Chambers et al. 2016), the MASTER global robotic network (Gorbovskoy et al. 2013), the All-Sky Automated Survey for Supernovae (Kochanek et al. 2017), the Gravitational wave Optical Transient Observatory (GOTO; Dyer et al. 2018), the Asteroid Terrestrial-impact Last Alert System (ATLAS; Tonry et al. 2018), and the Zwicky Transient Facility (ZTF; Bellm et al. 2019), OGLE-IV Udalski et al. (2015), and the upcoming Vera C. Rubin observatory (Ivezić et al. 2019). However, there is still a need for expanding and diversifying the global array of survey telescopes.

One reason is that some science cases require systems that are capable of accessing the entire sky at any given moment, while a typical ground-based observatory can access only ∼40% of the sky, with airmass below 2 during a clear night. Among such science cases are the study of gravitational-wave (GW) events and their electromagnetic counterparts (e.g., Abbott et al. 2017a, 2017b), the detection of rare and/or fast transients (e.g., Cenko et al. 2013), and the study of exoplanetary systems (e.g., Charbonneau et al. 2000; Nutzman & Charbonneau 2008; Swift et al. 2015). These and other science cases, as well as weather constraints and day-night cycle, call for around-the-globe, wide-field, high-cadence survey telescopes.

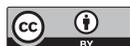

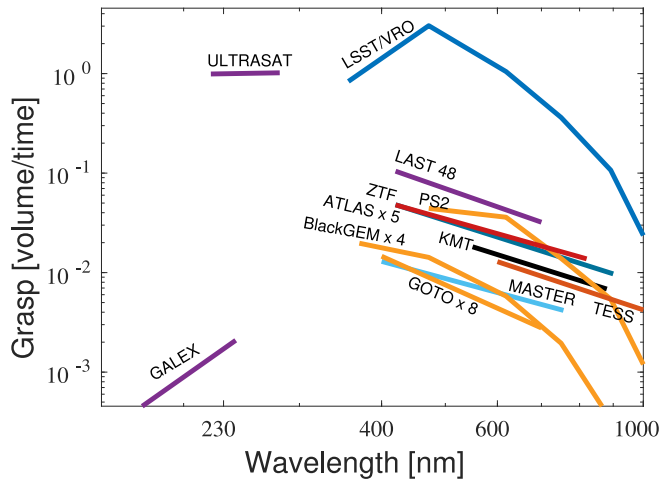

**Figure 1.** A comparison of LAST grasp, the amount of volume of space in which a standard candle is detectable per unit time, to leading on-going and soon to be commissioned sky surveys. Additional discussion is given in Ofek et al. (2023).

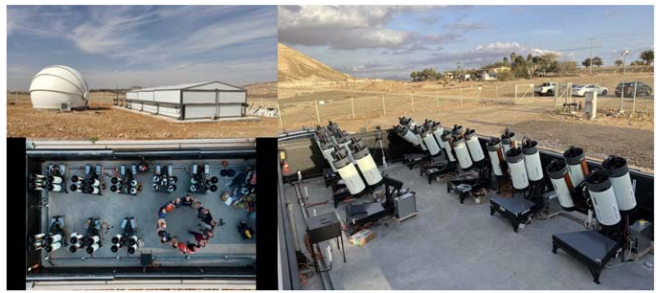

**Figure 2.** From left and clockwise: The LAST rolling roof enclosure; Panoramic view of the LAST 32 telescopes currently installed. The remaining 16 telescopes will be installed in summer 2023; Top view of the LAST enclosure.

One of the main obstacles when constructing a new sky survey facility is the high capital investment in hardware. Ofek & Ben-Ami (2020) argued for an approach in which the grasp, the amount of volume of space in which a standard candle is detectable per unit time, per cost can be increased significantly due to the availability of low-cost off-the-shelf (COTS) hardware, mainly large format, small pixel ($\lesssim 4\,\mu$m) CMOS detectors and fast optical tube assemblies with point-spread function (PSF) of $\sim 1''$. The Large Array Survey Telescope[8] (LAST) is a novel sky survey facility based on this approach. A comparison of the grasp of several leading on-going and soon to be commissioned surveys is shown in Figure 1, with additional discussion in Ofek et al. (2023). In terms of etendue, LAST and e.g., ZTF (Bellm et al. 2019) offer similar etendue. However, assuming that we are interested in the amount of volume of space in which a standard candle is detectable per unit time, we argue that grasp is the correct specification for comparison, which is the reason why we choose to use it throughout this and the accompanying papers, see Ofek & Ben-Ami (2020). It is the authors' hope that LAST at the Weizmann observatory in the south of Israel will be the first node of a global network based on its design and approach—that will deliver a global continuous coverage of the celestial sphere. We encourage interested parties to contact us in order to take full advantage of LAST capabilities. As the funds and manpower available for the operation and data analysis of such sky surveys is often the bottleneck for obtaining scientific results, collaboration with members of the community will determine how many of the scientific goals detailed in the following paper will be delivered.

In the following paper we describe the science objectives of LAST. We focus on science goals that will benefit the most from the three distinct advantages LAST has over existing sky surveys: high grasp; LAST geographic location—extending the global array of wide-field sky survey telescopes; and exposure times of $20 \times 20$ s with the temporal resolution required to resolve short-lived phenomena/effects. In order to increase the utility of the survey, coadd and calibrated object catalogs will be publicly released every 12 months, with transient events routinely reported to the TNS on a daily basis. Additional data products are considered, pending available funding. In Section 2 we give a short description of the LAST hardware and its capabilities. Section 3 discusses the survey strategy and cadence. Sections 4–6 describe the various science cases. Conclusions are given in Section 7. This paper is accompanied by two publications describing the technical aspects of the system (Ofek et al. 2023), and the data reduction pipeline (E. O. Ofek et al. 2023, in preparation; see also Ofek 2014, 2019; Soumagnac & Ofek 2018). Some of the plots in this paper were generated using tools available in Ofek (2014).

## 2. System Overview

LAST (E. O. Ofek and S. Ben-Ami, PIs), currently being commissioned at the Weizmann Astrophysical Observatory (WAO; Figure 2) in the south of Israel, is a cost-effective, wide-field survey facility. The ansatz of the survey telescope is that by combining many commercial, wide Field-of-View (FoV), COTS telescopes with small pixels ($\lesssim 5\,\mu$m) detectors, one can achieve a significantly higher survey grasp when compared to custom-built telescopes with larger aperture of similar monetary value (Ofek & Ben-Ami 2020). A more extensive LAST system overview is given in Ofek et al. (2023), and here we give a brief summary of the system design and capabilities. A LAST node has 48 telescopes mounted on 12 independent mounts and housed inside a rolling-roof enclosure. LAST telescopes are 27.9 cm F/2.2 Rowe-Ackermann Schmidt

---

[8] http://www.weizmann.ac.il/wao/

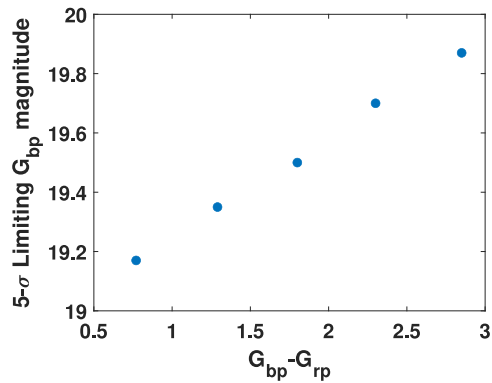

**Figure 3.** LAST measured limiting magnitude of a 20 s exposure with a single telescope, the Sony IMX-455 detector and no bandpass filter as a function of Gaia $G_{bp} - G_{rp}$ color measured at airmass ≈1. See Ofek et al. (2023) and the upcoming paper about LAST pipeline (E. O. Ofek et al. 2023, in preparation) for additional information.

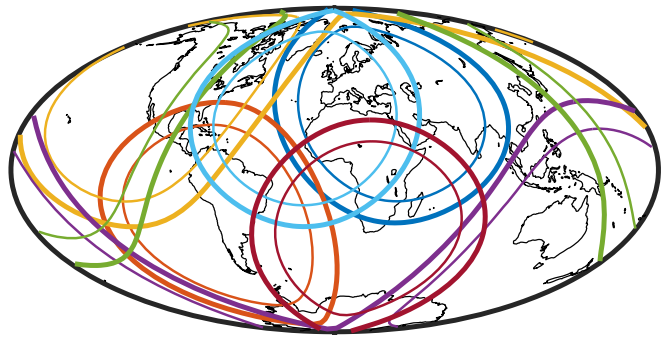

**Figure 4.** Projected cone of visibility with airmass contours of 1.5 (thin lines) and 2.0 (thick lines) from the geographic locations of several wide-field survey telescopes, plotted on the Earth map in Aitoff equal-area projection. Green represents Hawaii (e.g., Pan-STARRS1 and ATLAS), yellow for California (ZTF), red for Chile (e.g., DECAM and LSST), light blue for the Canary islands (e.g., GOTO, ATLAS), brown for South Africa (e.g., ATLAS), purple for Australia (e.g., Sky-Mapper), and dark blue for the first node of LAST located in the south of Israel. Rapid response to multi-messenger events requires additional wide-field telescopes in order to fill some of the missing gaps and to have better resilience to weather.

Astrographs (RASA) from Celestron, with a FoV of 7.4 deg² when coupled to full-frame (36 × 24 mm) detectors—QHY600M with a back-illuminated thermoelectrically cooled Sony IMX455 CMOS sensors. The use of commercially available backside-illuminated CMOS detectors with a price tag an order of magnitude lower than standard CCDs, will allow us to establish the limits of such detectors in a sky survey targeting versatile scientific goals, from the detection of extragalactic explosions to precision photometry of stellar hosts, when used in a setup that takes advantage of their low cost by building many copies of the same system. The typical 5σ limiting magnitude for point-source detection of a single telescope with the IMX-455 detector and no bandpass filter is 19.6 mag in a 20 s exposure (Gaia $G_{bp}$ band). When coadding 20 images, the number of exposures in a single visit, the limiting magnitude improves by 1.4 mag. The limiting magnitude as a function of the $G_{bp} - G_{rp}$ color is shown in Figure 3, see Ofek et al. (2023) and the upcoming paper about LAST pipeline (E. O. Ofek et al. 2023, in preparation) for a detailed discussion on filter relations and data calibration. Telescopes deliver an image quality of ∼2″.2–2″.8, which includes pixelation and the effect of seeing conditions at the site—see details below.

Each mount carries four telescopes, with the capability to either point all telescopes to a single FoV point ("narrow" mode), or to adjacent pointings, achieving a continuous large FoV ("wide" mode). Assuming the wide (default) configuration, the FoV for a single mount is ≈29 deg², with a full node covering ≈355 deg² or 0.8% of the celestial sphere in a single pointing (telescope footprints at a given time). Alternatively, in narrow mode, the system is equivalent in collecting power to a single 1.9 m telescope with a 7.4 deg² FoV. The modular structure of the system and the independent control of each mount allow for several intermediate modes between the wide and narrow field configurations.

LAST geographic location (30°2′54″N 35°1′31″E) allows us to fill-in part of the Asiatic gap in the global wide-field survey sky coverage eastward of the GOTO sky survey footprint (Dyer et al. 2018), see Figure 4, which gives the survey a significant benefit for some of the science cases described below. The WAO site has a very high fraction of clear nights: 93% of the nights have humidity levels of $RH < 70\%$, indicating the absence of low clouds throughout the year averaged over ten years[9]). While the presence of clouds at higher altitudes was not monitored routinely for a significant amount of time, our experience within the past year and a half suggest <20% overall cloud coverage. The median seeing at the site, measured over the course of a year, is ∼1″.35, with no apparent seasonal dependency. Therefore, seeing conditions are not expected to be a performance limiting factor for the vast majority of the observations.

Another topic to be investigated in the future is a novel calibration routine that will deliver the system transmission function along with the derived magnitude—and will make the introduction of e.g., bandpass and blocking filters o the optical train unnecessary. As this will require additional development from our team, we choose at the moment to report various system statistics in GAIA Gbp filter due to its precision and wide use across the community. The system pipeline, however, delivers cross calibration with various photometric filters (see E. O. Ofek et al. 2023, in preparation).

---

[9] Israel Meteorological Service (IMS) Neot Semadar Station https://ims.gov.il/en/node/242.

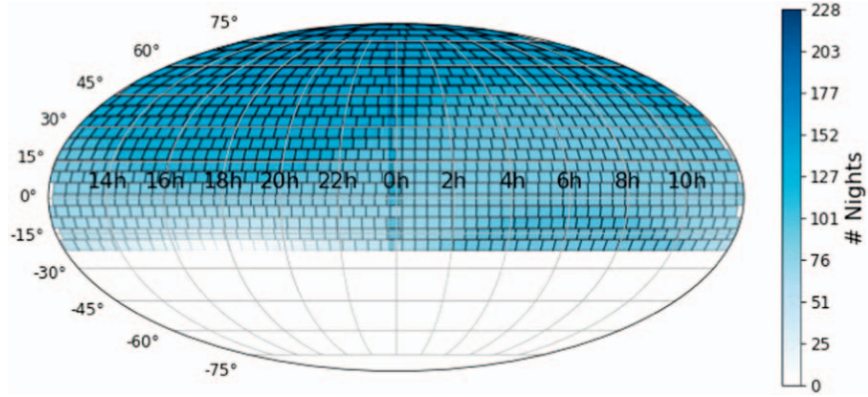

**Figure 5.** Number of nights within the calendric year 2023 in which a given field is observable for more than 5 hr at airmass <2, and with a lunar distance of >30 deg.

## 3. Cadence and Survey Strategy

The LAST modular design provides flexibility in the selected operation mode. In this publication, we base our estimates on two survey strategies using the wide-FoV configuration (i.e., each telescope is pointing to a different FoV, for a simultaneous sky coverage of $\approx 350\,\mathrm{deg}^2$): a high-cadence survey, and a low-cadence survey. We assume that 6 hr per night are dedicated, on average, to the sky surveys. The remaining observing time per night will be dedicated to follow-up observations, Target-of-Opportunity (ToO) observations, and additional programs that target specific science cases taking advantage of the LAST architecture. We emphasize that the strategy detailed below is preliminary and we expect to investigate various alternatives within the first year of operation.

### 3.1. High-cadence Survey

Eight mounts will conduct a high-cadence survey with an average of eight visits per night per pointing, where each visit is composed of $20 \times 20\,\mathrm{s}$ exposures (i.e., secondary cadence of $\sim 20\,\mathrm{s}$). Assuming each mount covers six fields on an average night, the footprint of the high-cadence survey is $\approx 1400\,\mathrm{deg}^2$. Within a given night, the fields observed in high cadence are required to have an airmass <2 for at least 5 hr. The average area of the sky per night fulfilling this condition is $\approx 10{,}400\,\mathrm{deg}^2$, varying between $\approx 6300\,\mathrm{deg}^2$ and $\approx 14{,}600\,\mathrm{deg}^2$ for the shortest and longest night of the year, respectively. Overall, 1240 pointings of individual mounts ($\approx 19{,}000\,\mathrm{deg}^2$) will follow these conditions in a calendric year, varying between footprints that fulfill the above condition for just a few nights to those that fulfill it for $\sim 6$ months—see Figure 5.

### 3.2. Low-cadence Survey

Four mounts will conduct a low cadence survey and complement the high-cadence survey described in the previous section with two visits per night for fields that fulfill the airmass <2 condition for at least 5 hr, but were not selected to be observed in the high-cadence survey. Each visit will be composed of $20 \times 20\,\mathrm{s}$. Assuming each mount covers 24 fields on an average night, the estimated footprint of the low cadence survey is $\approx 2750\,\mathrm{deg}^2$.

### 3.3. Polarization Survey

A subset of the LAST telescopes will be equipped with polarizing filters with four different alignments in the narrow-FoV mode. With $15 \times 60\,\mathrm{s}$ exposures in each field, a single LAST telescope can observe $3\pi$ sr in about 10 months of observing time. This will allow us to achieve a signal-to-noise ratio (S/N) of 100 (1000) for 20.1 (17.4) mag stars. This survey will allow us to better understand particle flows associated with astronomical objects as the polarization information (i.e., degree of polarization and polarization angle) are directly correlated with the magnetic field structure and its changes within the particle flows. Such flows, which are in many cases relativistic, are present in a variety of both extragalactic (e.g., Active galactic nuclei (AGNs), Gamma-ray bursts, Tidal disruption events, and Fast-Evolving Luminous Transients) and Galactic objects (e.g., X-ray binaries, supernovae, recurrent novae, and cataclysmic variable stars)—see Böttcher (2019), Reig et al. (2023), Krawczynski et al. (2022), Liodakis et al. (2023), Mandarakas et al. (2023); and V. Fallah Ramazani et al. (2023, in preparation) for further details.

### 3.4. Ground Support for ULTRASAT

The ULTRASAT space-borne telescope (Sagiv et al. 2014; Ben-Ami et al. 2022; Y. Shvartzvald et al. 2023, in preparation) is a 33 cm, $\approx 200\,\mathrm{deg}^2$ FoV space telescope operating at the near-ultraviolet bandpass of 230–290 nm, that is expected to be launched into geostationary orbit in 2026. ULTRASAT will reach a limiting magnitude of $\approx 22.5$ mag in $3 \times 300\,\mathrm{s}$ exposures for a blackbody target of $T_{\mathrm{eff}} \approx 20{,}000$ K, and

≈20.5 mag for an e.g., M4 dwarf ($T_{eff} \approx 3100$ K). In survey mode, ULTRASAT will continuously observe selected low-extinction and low-background fields for several months. This mode will benefit science cases such as the detection of shock breakout flares from supernovae (SNe), as well as characterization of stellar flares, in particular M-dwarf activity and variability. In ToO mode, ULTRASAT will be capable of rapid slewing to ∼50% of the sky at any given moment. With its large FoV and short wavelength observations, this mode is optimized for detection of electromagnetic counterparts of GW events detected by LIGO/VIRGO (e.g., Waxman et al. 2018; Abbott et al. 2020). LAST will monitor the ULTRASAT northern fields and will provide simultaneous data in the visible wave band, as the limiting magnitude of a visit co-add for LAST is similar to the limiting magnitude for e.g., cool dwarfs with ULTRASAT. For the latter sub field for which it achieves best image quality (i.e., an annulus at around 4° from the space borne telescope optical axis), we will consider utilizing 4 telescopes (i.e., a single mount) pointing to a single field to allow us to observe fainter sources. Such data are crucial to determine the time-dependent spectral energy distribution (SED) of the detected transients, providing information on, e.g., the effective temperature of blackbody sources, and the spectral shape of non-thermal sources (e.g., for stellar activity).

## 4. Extragalactic Astronomy

LAST magnitude limit and image quality allow us to target open questions in extragalactic astronomy. The large grasp of the survey offers an opportunity to detect rare events whose volumetric rate prevented detailed studies so far. In addition, the high-cadence survey described in Section 3.1 will allow us to investigate phenomena with short timescales and/or rapid evolution. A significant advantage of LAST over deeper sky surveys which utilize telescopes with larger apertures is that LAST grasp is driven by the large FoV rather than the depth. This makes follow-up observations straightforward even with small- and medium-sized telescopes, similar to existing surveys such as ATLAS (Tonry 2011) and NGTS (Wheatley et al. 2018). Among the topics in extragalactic astronomy targeted by LAST are the follow-up of GW and neutrino triggers in the visible band, the search for young SNe and tidal disruption events (TDEs), the study of lensed events, and uncovering the population of short transient events discovered in the past decade. These are described in details in this section.

### 4.1. Follow-up of Gravitational-wave Events

In 2015, the LIGO/VIRGO experiment detected the first GW event, resulting from the merger of two black holes (BHs; Abbott et al. 2016). Two years later, and after the discovery of additional black hole mergers, the collaboration announced the discovery of a neutron-star (NS)–NS merger event, GW170817 (Abbott et al. 2017a). About ten hours after the detection of the latter, ground-based observatories were able to detect the optical afterglow within the expected error volume (with a sky area of 40 deg², Abbott et al. 2017b). The multi-band photometric follow-up revealed low-mass ejecta expanding at $\gtrsim 0.2c$. This was deduced based on the expansion velocity of the blackbody radius, and the diffusion timescale, which is also suggestive of a relatively low opacity, $\kappa \lesssim 1$ cm² g⁻¹ ejecta (e.g., Rosswog et al. 2018; Waxman et al. 2018), as well as the width of the spectral features and radio observations. On a timescale of weeks, a radio and X-ray afterglow consistent with synchrotron emission due to interaction of the expanding ejecta with the interstellar medium was discovered (e.g., Abbott et al. 2017b), and VLBI observations suggested that the ejecta is jetted (Mooley et al. 2018).

NS–NS mergers may be used to measure the Hubble constant (e.g., Del Pozzo 2014), and can help answer several related questions. These include the nature of short gamma-ray bursts (GRBs; e.g., Nakar & Sari 2012), the nature of the observed gamma-ray emission from GW170817, and constraining the equation of state of nuclear matter. Early detection of the optical counterparts of GW events within an hour after the merger event is highly desired. Early UV-optical observations of afterglows in the first hours after the merger provide a strong test of theoretical models (e.g., Arcavi 2018) and have the potential to provide a measurement of the velocity distribution and the opacity of the ejecta (Waxman et al. 2018). Testing whether the optical-UV properties of all merger afterglows are similar to those of the one prototype detected so far would indicate how standard these events are—a substantial clue to their nature.

Given the large error regions expected from GW detectors (e.g., Nissanke et al. 2013), there are two strategies to identify optical afterglows. The first is to identify all known galaxies within the error volume and observe them according to some priority (e.g., galaxy mass, star formation rate). The second is to observe the entire error region. GW170817 was detected using both methods independently. However, GW170817 was detected at a distance of 40 Mpc and had an error region of only 40 deg². At such short distances, the local Universe is well known (e.g., Fremling et al. 2020) and the number of galaxies that require follow-up is relatively small. With the increased sensitivity of GW detectors, the typical distance of detected events will grow, and the efficiency of the first method will drop, as it becomes progressively more challenging to execute due to lack of complete galaxy catalogs and the significantly large number of galaxies within the given error regions. Therefore, we speculate that the scientific community will strongly rely on wide-field sky surveys that are able to fully cover these regions. Following this possibility, and in order to detect the optical afterglows shortly after a GW trigger (within an hour), we need observatories all around the globe, with LAST strategically located to partly fill the Asiatic gap, see Figure 4.

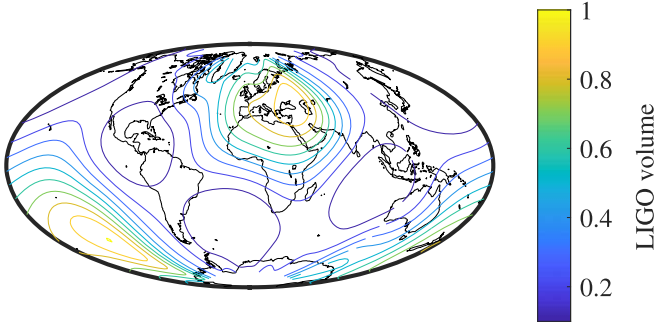

**Figure 6.** LIGO detector's volumetric response (arbitrary units) plotted against the Earth map in Aitoff equal area projection. The response map is shifted by 8 hr after the LIGO trigger to account for the approximate (expected) maximum light of the NS–NS merger event.

LAST limiting magnitude is sufficient to detect GW170817-like events at distances larger than the LIGO O4 horizon (190 Mpc; Abbott et al. 2020). Since the LIGO/VIRGO volume sensitivity is highly non-isotropic, with an order of magnitude directional variation (see Figure 6), by carefully monitoring the rough direction in which LIGO is most sensitive (at any given moment), we can increase the probability of near-real-time detection of NS–NS mergers optical afterglows. This can be further extended to a blind search: detection of NS merger events without a GW-detector trigger. In such a case, a detection of an EM signal will lead to trigger sub-threshold LIGO detections (e.g., Waxman et al. 2018). The maximum sensitivity of LIGO (which is currently the dominant instrument in terms of sensitivity) is roughly above North America and Australia (e.g., Abbott et al. 2017b, 2020). However, the maximum visible-light luminosity of NS–NS merger events is expected to take place a few to ∼24 hr after the merger. For an 8 hr delay between the merger and maximum light, the maximum volumetric sensitivity of LIGO will occur above the Middle East (see Figure 6). From Abbott et al. (2020), we estimate $\mathcal{O}(1\text{--}10)$ detections of NS–NS afterglows during the O4 run. For blind searches, we use Waxman et al. (2018), and estimate $\mathcal{O}(1)$ detections from each survey (i.e., the high cadence and the low cadence). This is comparable to the yield from the Vera Rubin observatory, with LAST expected to detect younger and brighter events.

### 4.2. Follow-up of High-energy Neutrino Events

High-energy neutrinos are produced by the interaction of cosmic rays (mainly protons, but also heavier nuclei) with ambient matter and photon fields (see e.g., Kurahashi et al. 2022, for a review). Neutrinos interact only weakly and are not deflected by magnetic fields. Therefore, they directly trace sites of hadronic acceleration and they can escape from dense environments that photons cannot leave.

Ten years after the discovery of a diffuse high-energy neutrino flux by the IceCube Observatory (Aartsen et al. 2013, 2015), the sources of those neutrinos are still largely unknown. A promising way to identify the neutrino sources is by detecting their electromagnetic counterpart. The IceCube Collaboration operates a real-time system, which releases the direction of likely astrophysical neutrino events through the GCN system with a median latency of 30 s (Aartsen et al. 2017). Several spatial coincidence between a single high-energy neutriono and objects of interest were claimed (Aartsen et al. 2018; Stein et al. 2021; van Velzen et al. 2021; Reusch et al. 2022).

Currently, IceCube issues ∼30 high-energy neutrino alerts per year (Blaufuss et al. 2019) and LAST can initiate rapid follow-up observations for ∼5 and observations within one day for ∼20 of them, as IceCube is most sensitive to neutrino events near the equator. LAST large FoV matches the typical angular resolution of high-energy neutrino events of a few square degrees. Due to its geographic location, LAST can provide follow-up observations for alerts that are not immediately observable by ZTF, PanSTARRS, DECAM, or LSST (see Section 2).

Rapid follow-up observations are crucial to detect short-lived electromagnetic counterparts such as FBOTs (Fang et al. 2019) or GRBs (Kimura 2022), including low-luminosity or orphan GRBs. Moreover, quick follow-up observations can constrain the explosion time of SNe and allow us to test whether a neutrino signal coincides with the core collapse as expected for choked-jet SNe (Chang et al. 2022). More neutrino alerts will be released by future neutrino observatories, such as KM3NeT (Margiotta 2014) and Baikal-GVD (Avrorin et al. 2018).

In addition, LAST will help to provide more complete catalogs of transient or variable neutrino source candidates such as certain types of SNe, TDEs, and AGN flares and the well-sampled light curves can be used for archival searches (see e.g., Stein 2019).

### 4.3. Early Detection of Supernovae

The increase in the number of sky surveys and survey grasp in the past twenty years resulted in an increase in the number of detected SNe from tens to thousands per year (e.g., Perley et al. 2020; Neumann et al. 2023). With this increase, our comprehension of the underlying physics has increased as well. However, there are still large gaps in our grasp of these cosmic explosions. Our understanding of the explosion mechanism is still incomplete, and models that adopt suggested mechanisms, such as the neutrino mechanism, fail to generate the observed kinetic energy of such events. Additional detections of young events, merely hours from the time of the first photons escaping from the explosion, have great potential in revealing the necessary data to close the gap in our understanding of these events. LAST, with its wide-FoV and depth, can yield these crucial observations. While LAST grasp offers only a modest increase compared to other surveys, the high cadence in which the survey operates offers an advantage for the detection of young events at the cost of the number of detections overall. We emphasize that while in the following section we discuss LAST capabilities to detect young SNe, much of LAST strength lies in its geographic location, cadence and grasp which will allow us to e.g., examine image co-adds retrospectively based on alerts from other surveys.

### 4.3.1. Shock Breakout

The first electromagnetic signal arriving at a remote observer of a star exploding as a SN is a brief flash of light occurring when the explosion shock breaks out of the surface of the doomed star. This flare is often called the SN shock-breakout flare (e.g., Waxman & Katz 2017). The duration of the shock breakout phase is of order 0.5 hr for a supergiant star SN progenitor, and the breakout burst properties depend on the progenitor radius, the burst velocity and the density at and during the flare. Observations of the burst can thus be used to constrain these parameters, and in turn to constrain global properties of the ejecta. LAST high cadence is appropriate to detect a SN explosion during the shock breakout phase; however, as most of the energy is expected to come out in the UV, a fortunate occurrence of a nearby event within the LAST high-cadence footprint is needed for the system to be able to pick up the visible-light tail of such an emission. Assuming shock breakout from red supergiants peaks at an absolute magnitude of $-15$ (e.g., Nakar & Sari 2010; Waxman & Katz 2017), LAST can detect those to distances of about 25 Mpc (redshifts $z \sim 0.005$). The rate of nearby SNe up to this distance is but a handful per year, e.g., 12 Type II SNe, typically associated with supergiant progenitors, were reported to the Transient Name Server (TNS[10]) during 2020–2021. With the LAST footprint and high-cadence observations, we can therefor expect to detect an $\mathcal{O}(1)$ shock breakouts during three years of operations. We take the opportunity and emphasize that with ULTRASAT continuous cadence and UV sensitivity (Sagiv et al. 2014; Ben-Ami et al. 2022; Y. Shvartzvald et al. 2023, in preparation), the rate of solid detections of shock breakout flares is expected to be significantly higher.

### 4.3.2. Shock Cooling

Following the shock breakout flare, the hot ejecta expands and cool. The release of the energy stored in the expanding ejecta initially dominates the energetic output of the SN and is commonly referred to as the shock-cooling emission. This would be the dominant emission in massive star explosions for a period that can extend for hours to days, depending on the size of the progenitor star and the effects of other energy sources, such as radioactivity from freshly-synthesized elements, or interaction with circumstellar material (CSM).

The shock cooling emission has been studied extensively. Theoretical studies suggest the cooling phase properties depend on the explosion and progenitor properties (e.g., Nakar & Sari 2010; Rabinak & Waxman 2011). As demonstrated by, e.g., ZTF, coupled to rapid UV follow-up with UVOT on board Swift (Gal-Yam et al. 2011), observations of infant SNe, discovered shortly after explosion, can thus yield valuable information about the explosion and progenitor properties (e.g., Nakar & Sari 2010; Rabinak & Waxman 2011; Soumagnac et al. 2020; Bruch et al. 2021). While LAST does not necessarily offers an advantage for events with shock cooling phases of several days when compared to e.g., ZTF, it does offer a clear advantage for events with shock cooling phases of hours due to its high grasp and observing strategy. We thus expect LAST to provide $\mathcal{O}(10)$ infant SN detections per year (similar to ZTF; Bruch et al. 2021) with significant representation for explosions with short cooling phases. This will have a significant contribution to this growing field by providing samples of events for which progenitor properties can be constrained. This will be further boosted in the near future when operating in tandem with ULTRASAT (see Section 3.4 above).

### 4.3.3. Flash-ionization Spectroscopic Observations

The very early discovery of SNe allows one to obtain spectroscopic observations within a day or so from the explosion, and probe the progenitor's immediate environment during the explosion. With the advent of adequate models in the future, this can determine CSM metallicity and mass, and obtain evidence for precursor ejections just before the SN explosion. Such observations (Gal-Yam et al. 2014; Khazov et al. 2016; Yaron et al. 2017; Bruch et al. 2021) revealed that a large fraction of SN progenitors are presumably embedded within a compact distribution of CSM, manifesting via transient high-ionization emission lines ("flash spectroscopy"). Several open questions remain, among them is the physical mechanism responsible for releasing this CSM, and the relation of this CSM ejection to higher mass counterparts and SNe precursors (e.g., Ofek et al. 2014; Strotjohann et al. 2021). Another important question is related to the optical depth of the CSM. Specifically, if its optical depth is large (e.g., larger than a few), then this CSM will modify the shock cooling emission and hinder our ability to use the shock cooling to estimate the progenitor radius and energetics. Therefore, better constraints on the CSM properties will, in turn, improve our estimates of the progenitor properties.

LAST high-cadence, wide-FoV survey will provide a large number of suitable SNe to be investigated in this manner. LAST, when coupled to a high-throughput transient classification machine such as the forthcoming Son-of-XShooter (SoXS) instrument (Rubin et al. 2018; Schipani et al. 2018), is especially promising as the emission lines detected in flash spectra presumably provide a direct probe of the surface composition of SN progenitors prior to explosion, e.g., a supergiant progenitor that has not experienced significant mass loss will exhibit solar metallicity through the observed CSM lines. For progenitors with significant mass loss, we expect to observe enhancement in carbon, nitrogen, and oxygen (CNO) products from the inner layers, see Gal-Yam et al. (2014). A caveat we would like to emphasize is the challenge of modeling the expected emission under non-local thermal equilibrium conditions (Dessart & Hillier 2011). Overall, this method could be highly informative to form a robust mapping between the progenitor and the SN properties.

---
[10] https://www.wis-tns.org/

### 4.4. Type Ia Supernovae

Type Ia Supernovae (SNe Ia) are standard distance indicators, which led to the extensive use of these explosive transients to measure the accelerated expansion of the Universe. The advent of wide-FoV sky surveys and the availability of extensive samples of SNe Ia have further increased the reliability of these cosmological probes (Betoule et al. 2014; Scolnic et al. 2018; Brout et al. 2019).

The current strategy for the use of SNe Ia as cosmological probes is based on relative distances, in which low-redshift events ($z \lesssim 0.1$) are used to anchor events at higher redshifts. In addition, low-$z$ SNe Ia are used to study large-scale structure in the local Universe. Until recently, low-redshift samples used in establishing relative distances have been heterogeneous, leading to significant systematic uncertainties (Dhawan et al. 2021). Modern, untargeted, wide-FoV surveys have increased our discovery rate of these explosive transients by an order of magnitude (e.g., see Tonry 2011; Tonry et al. 2018; Graham et al. 2019; Jones et al. 2021), and several leading surveys have now gathered extensive uniform samples of low-$z$ SNe Ia (e.g., Dhawan et al. 2021) making it possible to overcome the aforementioned uncertainties.

We estimate the number of events that LAST will detect using ZTF detection rates (1200 SNe Ia per year; Dhawan et al. 2021). With a similar limiting magnitude per visit (30 s exposure for ZTF versus $20 \times 20$ s exposures with LAST), the ZTF high-cadence survey is similar to LAST low-cadence one (e.g., Kupfer et al. 2021), and we expect to discover a few hundred SNe Ia each year. LAST high-cadence survey is expected to deliver an additional $\mathcal{O}(10^2)$ events. The wide-FoV of LAST will allow us to detect young, low-redshift events for which follow-up spectroscopic observations can be easily obtained early on, with a significant number of events at redshifts for which host galaxy distances can be accurately measured. This is a property of systems for which grasp is dominated by area rather than depths, such as ATLAS (Tonry et al. 2018).

### 4.5. Super-luminous Supernovae

Superluminous SNe (SLSNe) are explosions in which the peak luminosity far exceeds those of normal SN explosions, often by at least two orders of magnitudes (Gal-Yam 2019). This class of catastrophic explosions has been identified only in the past two decades, and the nature and physical mechanism driving these explosions are still not understood (Richardson et al. 2002; Ofek et al. 2007; Quimby et al. 2007, 2011; Gal-Yam et al. 2009; Cartier et al. 2022). With a limiting magnitude of $\sim 22$ mag in nightly 8-visit coadds, and reaching 23 mag in weekly coadds, LAST is sensitive out to redshifts of $z \sim 0.6$ and $z \sim 0.9$, respectively, for a fiducial event with a peak magnitude of $-21$ mag. Since these events are slowly-evolving, with typical rise times in excess of 30 days (Gal-Yam 2019), LAST surveys would provide a large sample of such events. Deeper stacks can be used to select high-redshift events that are easily separated from other SN interlopers by their slow rise. Scaling from the ongoing ZTF survey (e.g., Perley et al. 2021), several events per week at low-redshift, and a similar number of events at high redshift, are expected. Such a sample would provide a data set to study the properties of SLSNe and in particular probe dependencies on redshift.

A subset of SLSNe display luminous initial "bumps" prior to the main peak (e.g., Leloudas et al. 2012; Nicholl et al. 2015; Vreeswijk et al. 2017; Gal-Yam 2019). LAST continuous monitoring will be especially useful in finding these events, and will enable spectroscopic coverage during this initial bump, possibly shedding light on the physics driving these preliminary peaks and the explosion mechanism of SLSNe in general.

### 4.6. Short-duration Transients

With the increase in survey telescope grasp and cadence, sky surveys are discovering have discovered transient events with a timescale of several days, e.g., PTF 09uj (Ofek et al. 2010; Drout et al. 2014; Ho et al. 2018), SN2018cow (Perley et al. 2019; Ho et al. 2023), and AT2018lqh (Ofek et al. 2021). There are still several astrophysical phenomena that withhold a secure detection. Among these are accretion-induced collapse events (e.g., Dessart et al. 2006; Sharon & Kushnir 2020) and orphan GRBs (Levinson et al. 2002; Nakar et al. 2002; Gal-Yam et al. 2006).

The nature of most of these events is uncertain, and it is possible that there is a diversity of physical mechanisms that are responsible for these transients. For example, AT 2018lqh was suggested to be powered by a low-mass ejecta, of the order of $O(10^{-2})$ $M_{\odot}$, composed mostly from radioactive material (e.g., Ni$^{56}$), while AT 2018cow is likely powered by the interaction of the ejecta with the CSM. Interestingly, such interaction-powered transients are promising candidates for the acceleration of the highest-energy Galactic cosmic rays (Katz et al. 2011; Murase et al. 2014). FBOTs and interaction-powered SNe in the local universe identified with LAST will be followed up by current and next-generation gamma-ray telescopes such as H.E.S.S. or CTAO. FBOTs are all frequently detected in the optical band and also promising neutrino sources (Fang et al. 2019).

Understanding the nature of such events requires large samples and spectroscopic and multi-wavelength follow-up. With the high cadence survey, and to a smaller extent with the low cadence survey, LAST high grasp and optimization with respect to FoV rather than depth will allow us to discover such events and conduct follow-up spectroscopic observations. Using rate estimates from Ho et al. (2018) for transients with peak magnitude $m = 18$ which fade by two magnitudes in less than three hours, we can expect $\mathcal{O}(100)$ events per year to be detected by LAST.

### 4.7. Gamma-Ray Bursts

Dedicated fast-response facilities have increased our understanding of GRBs. The questions that are likely to benefit the most from the detection and observations of additional events are: (i) the existence of a dirty fireball, or geometric effects that produce optical afterglows with no prompt $\gamma$-rays emission (e.g., Rhoads 2003); (ii) the relation of short-duration GRBs to NS–NS mergers (e.g., Berger 2014); and (iii) the energy range of the peak flux and its extension to lower energies below $E_{\rm peak} \sim 100$ KeV (e.g., Ghirlanda et al. 2009). LAST can provide effective ways to study GRBs and find clues to these questions.

#### 4.7.1. Orphan GRB Afterglows and Blind Searches

Several predictions suggest that GRBs may produce optical afterglows with no prompt $\gamma$-ray emission. These include viewing-angle effects (Rhoads 1997; Perna & Loeb 1998; Nakar et al. 2002), as well as processes that suppress the high-energy emission (Dermer et al. 2000; Rhoads 2003). One way to search for such events is at longer wavelengths (e.g., radio). To this date, these searches produced only a handful of candidates (e.g., Levinson et al. 2002; Gal-Yam et al. 2006; Ofek 2017; Law et al. 2018). At shorter wavelengths, current limits obtained by ZTF suggest that the fraction of events without $\gamma$-ray emission cannot be significantly higher than the rates of classical GRBs (610 yr$^{-1}$, Cenko et al. 2015; Ho et al. 2020). In order to better define the actual rate, a larger sample is required.

We use results from Cenko et al. (2009) to estimate the number of GRBs LAST can detect through blind searches. Out of 28 Swift GRBs that were promptly observed using the Palomar 60 inch telescope (Cenko et al. 2006), 20 showed visible wavelength afterglows brighter than magnitude 20.5 within a few hours following the GRB. The all-sky rate of Swift GRBs out to $z = 3$ has been estimated to be $1455^{+80}_{-112}$ yr$^{-1}$ (Lien et al. 2014). We conclude that the expected yearly GRB afterglow rate for LAST high cadence survey is

$$R_{\rm GRB} \approx 8 \frac{\Delta T}{5 \text{ hr}} \frac{\Omega_{\rm fast}}{1400 \text{ deg}^2} \text{ yr}^{-1}, \quad (1)$$

where $\Omega_{\rm fast}$ is LAST high cadence footprint, and $\Delta T$ is the number of hours dedicated to observe this footprint per night.

One question is how to identify such events? A pro-active approach is to point the LAST telescopes simultaneously to a subset of the FoVs observed by Swift BAT (the latter has a FoV of 2 sr–∼10× LAST FoV, Gehrels et al. 2004) in order to observe the rise, or the absence of the GRB afterglow in real time. Regardless of whether this approach is feasible at a specific time, we expect these events to evolve very fast, and to be often accompanied by late-time radio afterglows. Hence, follow-up observations triggered immediately after detection will deliver additional clues to the nature of the candidates (e.g., Cenko et al. 2013). Additional detailed photometric and spectroscopic observations for selected high-quality and high-probability candidates will allow us to conduct in-depth studies of nascent events with and without $\gamma$-ray emission, and shed light on the physics driving the observed event.

#### 4.7.2. ToO Observations of GRBs

The typical Fermi-GBM error region has an area of hundreds of square degrees and suffers from a systematic bias (Goldstein et al. 2020). LAST FoV is large enough for searching such afterglows promptly over the entire error region, and its location within the Asiatic gap holds a unique advantage in this case. Specifically, about 20% of the GRBs are short-duration and hence believed to be related to NS–NS mergers. However, the number of short-duration GRBs with known afterglows and host galaxies is small (e.g., Fong et al. 2015, 2016). Therefore, increasing this sample is of great interest in order to study the possible relation of this population to NS–NS mergers.

#### 4.7.3. Prompt Optical Flash from GRBs

In several cases luminous prompt optical flashes are observed from GRBs (e.g., GRB 990123; Feroci et al. 1999). The leading explanation for this phenomena is emission from the reverse shock (e.g., Waxman & Draine 2000). Even for under-luminous GRBs ($E \sim 10^{51}$ erg) at high redshifts, these are expected to be brighter than magnitude 18. LAST coverage of 0.8% of the celestial sphere to 19.6 mag on a 20 s timescale will allow us to detect such events up to very high redshifts. If such optical flashes accompany all GRBs (see however, e.g., Park et al. 1997), LAST is expected to find about 3 events per year.

### 4.8. Tidal Disruption Events

When a star passes sufficiently close to a supermassive black hole (SMBH), it is torn apart by tidal forces. For certain stars and BH masses ($\lesssim 10^8 M_\odot$ for a non-rotating SMBH and a G-dwarf), the disruption (called a tidal disruption event, or TDE) will occur outside of the event horizon and will be accompanied by an observable flare (Hills 1975; Rees 1988).

Initially, the flare was thought to be solely powered by the accretion of the bound stellar material onto the BH, and thus was expected to peak in the X-rays. Arcavi et al. (2014) argues for a new optical-UV class of TDEs (see van Velzen et al. 2020; Gezari 2021, for a review). The events in this class exhibit blue colors, peak absolute magnitudes of $\sim-20$ and a $\sim t^{-5/3}$ decay, which has been suggested as a signature of TDE emission (Rees 1988; Evans & Kochanek 1989; Phinney 1989).

The source of the optical-UV emission in these events is debated: Is it from the outer shocks of the leading stellar debris hitting the debris tail due to general-relativistic precession, before the accretion disk is formed (e.g., Piran et al. 2015)? Is it re-processed accretion-disk emission (e.g., Guillochon et al. 2014; Roth et al. 2016)? Or is it a combination of both? Answering these questions requires detailed observations of many events.

The strategy for finding TDEs with LAST will be similar to that outlined for SLSNe, since both types of events have similar luminosities and timescales. Scaling from ZTF, LAST is expected to find roughly 10 TDEs that are brighter than 20th magnitude and hence accessible for spectroscopic follow up per year (TDEs are intrinsically rare, with a rate of $\sim 10^{-4}$–$10^{-5}$ events per galaxy per year Wang & Merritt 2004; Stone & Metzger 2016). Spectroscopic follow-up is crucial for TDE classification as most transients in galaxy centers are not TDEs (even when focusing on quiescent galaxies; Arcavi et al. 2022), and so in this case as well, coupling to a high-throughput transient classification machine, such as the forthcoming SoXS instrument (Rubin et al. 2018; Schipani et al. 2018), is especially promising.

As part of either the high-cadence or low-cadence survey, LAST will also provide nightly monitoring of TDEs. Recently, we discovered that at least some TDEs display changes in their light curves on daily timescales (S. Faris et al. 2023, in preparation). Such behavior is a challenge for current TDE emission models to reproduce, indicating that they might be missing some important physics. However, we are not yet certain how common such a behavior is. LAST will be able to quantify this behavior for all TDEs in the northern sky.

The emission from the known optical-UV class of TDEs peaks in the UV, making them prime targets for ULTRASAT. A persistent UV detection from ULTRASAT will help rule out most transient contaminants (such as SNe, which fade in the UV very quickly), thus allowing the identification of TDEs out to higher redshifts, without requiring spectroscopic vetting. LAST will be able to complement the ULTRASAT data with optical photometry, when stacking data on a daily and weekly basis.

A yet undetected subclass of TDE is the disruption of a white dwarf (WD). Such an event would constitute a smoking gun evidence for the existence of intermediate-mass black holes (IMBHs), as only they would disrupt a WD outside of their event horizon, producing an observable flare. Such TDEs would be visible on much shorter timescales than the SMBH disruptions seen so far. The high cadence of LAST will thus also be useful for alerting on rapidly-evolving TDEs as possible indicators of IMBHs.

### 4.9. Gravitational Lensing

#### 4.9.1. Lensed Quasars

Lensed quasars allow us to probe the dark matter content of galaxies (e.g., Maoz & Rix 1993), measure galaxy mass evolution (Ofek et al. 2003), study the magnified hosts of quasars, and measure cosmological parameters such as the Hubble constant (Saha et al. 2006; Oguri 2007; Birrer & Treu 2021; see however some important limitations in Kochanek 2002; Blum et al. 2020; Kochanek 2020).

The typical image separation of lensed quasars is of the order of $0.''5$, just below the seeing disk induced by Earth's atmosphere. Therefore, detection of these objects and measuring time-delays between multiple images is challenging. Springer & Ofek (2021a) showed that the combined flux of a lensed quasars and its center-of-light (photo-center) astrometric position time series, are rough analogs to observing the resolved lensed images. As the LC flux is expected to change by a few percent in a timescale of $\sim 10$ s days and the photocenter position by $\sim 20$–50 mas in a similar time span (e.g., Tewes et al. 2013; Tubín-Arenas et al. 2023), this method relies on high cadence wide-FoV sky surveys, allowing us to detect lensed quasars and measure their time delays without resolving the system. LAST photometric precision and measured astrometric accuracy of $\sim 50$ mas for co-add of 16 visits for magnitude 20 objects are sufficient for this purpose (see upcoming paper about LAST pipeline, E. O. Ofek et al. 2023, in preparation), and will allow us to test this method and potentially find lensed quasar, or measured the time delay of known lensed quasars, brighter than 20 mag.

#### 4.9.2. Lensed Type Ia Supernovae

In recent years, with the increase of synoptic sky surveys grasp, some surveys started finding lensed SNe Ia (e.g., Quimby 2014; Goobar et al. 2017). Similarly to quasars, lensed SNe Ia offer a method to study the galaxy-mass distribution and to measure the Hubble parameter (see also Section 4.9.1). Unlike quasars, the absolute magnitude of SNe Ia can be in principal estimated from their light-curve width (Phillips 1993), which gives additional information that can help remove the mass-sheet and similar degeneracies (e.g., Kochanek 2002, 2020; Blum et al. 2020). Mörtsell et al. (2020) estimate that given the uncertainty of SNe Ia's absolute magnitude, it may be possible to estimate the Hubble parameter to 5% accuracy. Goldstein & Nugent (2017) investigated the impact of microlensing on the accuracy of the measured time delays, and concluded it may have up to a few percent-level influence on the time-delay measurements.

Assuming the LAST low-cadence survey scans 2750 deg$^2$ and the daily coadd of the high cadence survey scans 1400 deg$^2$, the two surveys are expected to contribute roughly equally to the detection of lensed type Ia SNe. For example, LAST can detect a source with an absolute magnitude of $-18.5$ to $z=0.16$, and $z=0.26$, in the low cadence and daily-coadded high cadence, respectively. However, in the regime of low-redshift, the optical depth for lensing increases with redshift, and so daily coadd (or even a few-nights coadd) of the high cadence data may be more prolific for finding lensed SNe Ia.

An important question is how to identify such lensed SNe among normal SNe. Goldstein & Nugent (2017) suggest to search for SNe on top of elliptical galaxies, as the cross section for lensing is dominated by elliptical galaxies (Oguri & Marshall 2010). Since for such galaxies photometric redshift estimates are reliable, one has to look for SNe which are brighter than the typical SN Ia in that redshift. This method will be most effective when the lensing galaxy flux dominates the lensing event. An additional method that covers all lensing galaxies is presented by Springer & Ofek (2021a), who argue that by using the astrometric signal we can distinguish between lensed and non-lensed objects.

For the high cadence survey described in Section 3.1, and assuming that a photometric accuracy of 10% is required for the detection of lensed Type Ia SNe, we estimate a detection rate of $\mathcal{O}(1)$ lensed SN Ia per year based on Oguri & Marshall (2010), and assuming $\theta_{\rm E}$, the Einstein ring radius, is larger than $\approx 0\rlap{.}{''}1$, to guarantee precise astrometric registration as the event evolves.

### 4.10. Reverberation Mapping

The existence of SMBHs, less than 1 billion years after the time of recombination, is not well understood (e.g., Inayoshi et al. 2020). Furthermore, the observed scaling relations between SMBHs and their host galaxies (e.g., Magorrian et al. 1998; Ferrarese & Merritt 2000) likely tell us something fundamental about the growth of galaxy mass. Hence, measuring the mass of BHs in quasars is of great importance.

The most reliable method to measure, in large numbers, the mass of the BHs residing in quasars is the reverberation mapping technique (Peterson 1993; Kaspi et al. 2000, 2005). However, this method is observationally expensive, requiring either spectroscopic observations or narrow-band imaging. Springer & Ofek (2021b) presented a novel method that has the potential to measure the reverberation timescale of quasars using a single broadband synoptic observation. The method relies on the fact that the statistical model describing quasar light curves is roughly known (i.e., a power-law power spectrum). While the advantage is that reverberation mapping can be done using single-broadband observations, simulations suggest that this method likely requires a large number ($\gtrsim 500$) of photometric observations, preferably without large temporal gaps, and that the equivalent width of the broad emission lines will be a significant fraction (e.g., $\gtrsim 10\%$) of the effective bandwidth (Springer & Ofek 2021b). This method has the potential to deliver a large number of reverberation timescale measurements. LAST is not unique in that aspect and likely other surveys can be used for this task. However, given the importance of minimizing the gaps between observations, using data from several surveys to fill observational gaps may be important, and LAST unique location holds a significant advantage in this aspect.

### 4.11. AGN Variability

AGNs are a peculiar class of galaxies that host a SMBH ($\gtrsim 10^6 M_\odot$) at their center and are known to be the most powerful steady emitters in the Universe, with luminosities of the order of $10^{14} L_\odot$. They are powered by accretion of the surrounding environment of stars, gas and dust onto the SMBH. In the past decades, AGNs have been identified in a large variety of subclasses, depending on their physical and geometrical properties observed throughout the whole electromagnetic spectrum (see Padovani et al. 2017 for a recent review).

A fraction of all observed AGNs shows a relativistic jet of plasma that develops up to Mpc scales. A rare subclass of these jetted AGN ($\sim 1\%$ of all AGN) have this jet pointing at a small angle ($\leqslant 10$ deg) toward Earth and are classified as blazars. Blazars are observed from radio wavelength up to TeV gamma-rays, and their emission in the optical band is typically dominated by synchrotron radiation from relativistic electrons in the jet. Part of their optical emission can also be explained by a thermal contribution of the accretion disk, especially in Flat Spectrum Radio Quasars. Given their exceptional properties as cosmic accelerators, they have been suggested as possible neutrino source candidates, (e.g., Mannheim 1993; Atoyan & Dermer 2001), with several sources identified as potential neutrino blazars (e.g., Aartsen et al. 2018; Garrappa et al. 2019; Franckowiak et al. 2020).

A distinct characteristic of AGN is the variability of their emitted radiation observed at all wavelengths and in a wide range of timescales. In particular, blazars are among the most variable known extragalactic sources, showing peculiar behaviors at both intra-day and long-term timescales. The study of their flaring activity is crucial to estimating the intrinsic properties of the emission regions, the radiation processes as well as the geometrical effects of the jet structure and orientation (e.g., Bhatta 2021; Webb et al. 2021).

LAST high-cadence and low-cadence surveys will provide a nightly monitoring of $\mathcal{O}(10^4$–$10^5)$ AGNs, including a significant number of identified blazar candidates. Along with the photometric precision reached by LAST, it will be possible to study both long and short-term variability of these objects. The characterization of statistical properties of their temporal behavior (e.g., power spectral density, flux distributions and more) and the search for distinctive features like quasi-periodic oscillations will help shed light on the mechanisms behind this variability, that are still largely unclear (e.g., Jorstad et al. 2022; Yu et al. 2022; Pininti et al. 2023). Moreover, the relation between variability and the intrinsic physical properties of these sources can help with the identification of different sub-classes of AGN (e.g., López-Navas et al. 2023).

The joint observations from the polarization survey (Section 3.3 will also be important for the study of AGN magnetic field properties (e.g., Marscher & Jorstad 2021). Such broad monitoring of AGN population will be of fundamental importance in a multi-wavelength and multi-messenger context. The simultaneous observations of ULTRASAT northern fields as well as data from high-energy surveys like Fermi-LAT (Atwood et al. 2009) will give important insights on the processes behind the broadband emission observed from these powerful sources (e.g., Liodakis et al. 2019; de Jaeger et al. 2023).

## 5. Galactic Astronomy

In the following section we list some of the main questions in Galactic astronomy to be targeted by LAST within its first years of operation. These include the study of WDs and the search for minor bodies transiting these objects, the search for short-period variable objects, and the search for exoplanets transiting main-sequence (MS) stars. In addition, we discuss LAST potential as a microlensing experiment, and consider not only the first LAST node at the Weizmann Astrophysical Observatory, but also the possibility of constructing a LAST node in the southern hemisphere for a galactic plane microlensing event search.

### 5.1. Precision Photometry and Exoplanets

The majority of known exoplanets were discovered using the transit method (e.g., see review by Deeg & Alonso 2018). The transit depth of an Earth-sized planet orbiting an M-Dwarf (M4V), a prime target for exoplanet atmospheric studies, is ∼1 mmag. Currently several ground surveys can achieve this benchmark precision (e.g., Tregloan-Reed & Southworth 2013; Bayliss et al. 2022). Assuming an exoplanet host is observed by a single telescope, LAST is expected to obtain 1 mmag precision for magnitude $G_{\rm bp} = 13$ targets in a single exposure, or for $G_{\rm bp} = 14.5$ in a single visit which is 20 × 20 s exposures, see Figure 7 and Ofek et al. (2023). Therefore we will conduct a blind search for exoplanets transiting host stars with a magnitude of $G_{\rm bp} \sim 13$–15.5. In addition, a focused survey will target known planet hosts from TESS and other sky surveys and search for additional members of known transiting planetary systems benefiting from a longer temporal baseline. Compared to existing ground surveys, LAST offers a higher collecting power (e.g., a factor of 8 higher than NGTS; Wheatley et al. 2018), an increased number of independent apertures, and a larger FoV.

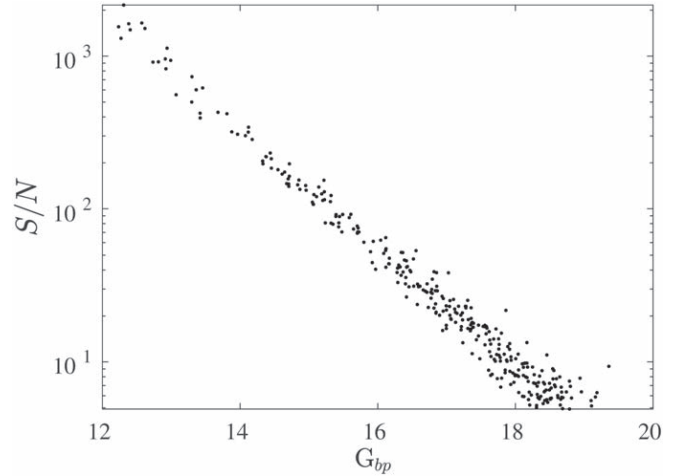

**Figure 7.** Measured S/N for bright objects derived from an exemplar 20 s exposure taken by LAST at the Weizmann observatory. The S/N was estimated with aperture photometry using a large photometric aperture of 6 pixels radius (see Ofek 2014; Ofek et al. 2023). Hence, it is sub-optimal for faint objects. The scattering in S/N shown for bright sources is most likely due to the lack of a bandpass filter in the optical train and the different SED between sources in the image.

One of LAST goals is to study the possibility to perform simultaneous precision photometry for thousands of stars at a level better than 1 mmag using ground-based telescopes. A possible approach to this challenge is to utilize LAST large number of independent telescopes to observe a single field. This has the advantage that any independent source of noise in the system, such as flat-field errors and scintillation noise, can be reduced further by the number of telescopes, by $\sqrt{N_{\rm tel}}$. We emphasize that the use of CMOS detectors in LAST add an additional challenge, as systematic errors will be introduced due to the nonlinearity of these detectors, dead and hot pixels, and the variable PSF inherent in the system.

Another unique opportunity offered by LAST is the use of multiple filters simultaneously on independent apertures observing the same FoV. In principle, this will allow us to study the chromatic behavior of both instrumental and atmospheric noise as well as their dependence on the stellar

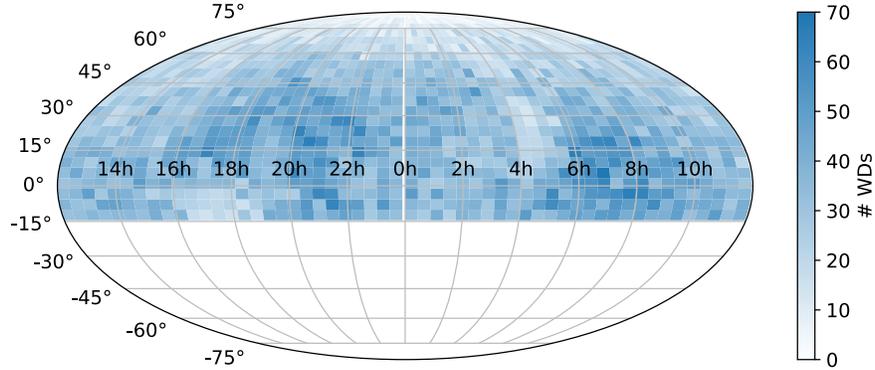

**Figure 8.** The WD distribution within the high cadence footprints. Each square represents the FoV of four telescopes on a single mount in the wide-FoV mode. Here we include WD candidates that are within the limiting magnitude shown in Figure 3, and with $P_{WD} > 0.95$ in the Gaia EDR3 WD catalog (Gentile Fusillo et al. 2021).

SED, and potentially remove them using this chromatic dependency. Following the removal of chromatic-dependent noise, we can study the spatio-temporal correlations of atmospheric transmission, in order to remove low-opacity clouds using many stars in the FoV.

### 5.2. White Dwarfs: Binaries, Transits, and Pulsations

The continuous series of $20 \times 20$ s exposures per visit, combined with the high cadence survey and LAST large FoV will allow us to monitor $\mathcal{O}(10^3)$ WDs simultaneously on short timescales. This will open new grounds in the search for WD binary systems, the study of $g$-mode pulsations, and the search for planets and planetesimals orbiting WDs. We find ∼40,000 WDs (high-probability WD candidates, $P_{WD} > 0.95$) within the limiting magnitude curve shown in Figure 3 from the Gaia EDR3 data (Gentile Fusillo et al. 2021) in the LAST high cadence footprint. The average number of WDs in a given exposure of 355 square degrees is ∼350, see Figure 8. Further details are given below.

#### 5.2.1. Binaries

LAST planned observing strategy has several advantages for detecting WD binary systems. Currently, only ≈130 double WD systems with periods ⩽1 day are known, ≈15 of them were discovered through the eclipse method (Korol et al. 2022). The combination of $20 \times 20$ s exposures per visit and a high cadence constitutes a significant advantage for the detection of short-period eclipsing binaries, with the emphasis that an efficient campaign will require developing the methodology to reject false positives as the transit duration of such systems is expected to be of the order of $\mathcal{O}(1)$ minute, and might require observing the same field with more than one telescope, not the least to guarantee a large sample of WDs is explored. Specifically, we will aim for mapping the population of eclipsing short-period double WDs. Double-degenerate short-period binaries, with periods ranging from minutes to several hours, are potential GW sources that can be targeted by future space-based GW missions (e.g., Amaro-Seoane et al. 2017; Korol et al. 2022). Given the capability to monitor orbital decay through observations in the electromagnetic spectrum, these targets are expected to be useful verification and calibration sources (Kupfer et al. 2018). These systems are also potential SN Ia progenitor systems through the double-degenerate channel (e.g., Webbink 1984; Toonen et al. 2012; Maoz et al. 2014). Thus, the improved statistics of these systems, with well studied orbital parameters, will allow us to estimate what fraction, if any, of SNe Ia originates from this channel (see however, Dong et al. 2015). Assuming the fraction of double WDs with separations of ≲0.25 au follows estimates for the fraction of double WDs with a separation below ≈4 au ($f \approx 10\%$; Maoz et al. 2018), we estimate a potential of $\mathcal{O}(10^1$–$10^2)$ eclipsing double WDs systems can be discovered by LAST in 2–3 yr in a dedicated analysis.

Additional binary populations we will investigate are eclipsing WD+M dwarfs and WD+brown dwarfs. The former are presumably the most frequent end-state of the common-envelope phase. They provide insights to many aspects of binary evolution, such as the impact of increased stellar rotation on flare rates, disk disruption, tidal effects, angular momentum exchange (Morgan et al. 2016), age determination of field M-dwarfs, and chromospheric activity-age relation calibration (Silvestri et al. 2005). The latter are candidates for cataclysmic variables (CVs) with sub-stellar donors, of which only ∼10 are currently known (e.g., French et al. 2023). These systems are useful laboratories for studying the transition from a post common-envelope binary to a pre-CV (e.g., Longstaff et al. 2019). In addition, the detection of new detached WD+brown dwarf binary systems will add to the small number of benchmark systems (e.g., van Roestel et al. 2021) which will allow us to secure the brown-dwarf mass–radius relation. Discriminating between type of binaries will be based on photometric and spectroscopic follow-up observations, for example, with the upcoming SoXS UV-VIS-NIR spectrograph (Rubin et al. 2018; Schipani et al. 2018).

### 5.2.2. Transits: Exoplanets and Debris

All but a few of the known exoplanets orbit MS stars, which will eventually evolve into WDs. It is unlikely that planets in close orbits (below ∼2 au) will survive this transition, as their host star climbs the red-giant branch, and its envelope expands to more than 1 au. However, due to gravitational interactions as the system becomes unstable, planets further out can migrate into closer orbits after the host's transition into a WD is completed (Debes & Sigurdsson 2002). Planets can also form out of gas near the WD via the interaction or merger of binary stars (i.e., second-generation planets; Livio et al. 2005). It was proposed that the first confirmed exoplanets, discovered orbiting a NS, have been formed in a similar manner from a disk created after the SN explosion (e.g., Wolszczan & Frail 1992).

There is accumulating evidence that the remains of old planetary systems are very common around WDs (see Veras 2021, for a recent review). These are manifested as metal pollution from accreted planetary material in WD atmospheres (e.g., Koester et al. 2014), spectral emission or absorption lines from circumstellar gas disks (e.g., Manser et al. 2020), infrared excess from dusty debris disks (e.g., Swan et al. 2020), and transits of planetary debris. The first transiting planetary debris orbiting a WD was discovered in Kepler K2 data (Vanderburg et al. 2015). In the past couple of years, this number has grown rapidly, especially since the first data release of ZTF (Vanderbosch et al. 2020; Guidry et al. 2021). Currently, ∼10 WD systems with transiting debris are known (see Farihi et al. 2022, and references therein), along with a growing number of candidate systems. While some of these systems show transits with orbital periods of a few hours (within the tidal radius for rocky material), others show no short timescale periodicity and could have wide eccentric orbits.

Recently, the first transiting planet orbiting a WD was discovered using TESS data, with an orbital period of ≈1.4 days (Vanderburg et al. 2020). A transit observed by a single LAST telescope during a test run in 2020 September 14 is shown in Figure 9, demonstrating our capabilities to observe such objects. A dedicated survey for transiting exoplanets and minor bodies around WDs is thus both timely and critical for advancing our understanding of these systems. The study of WDs polluted by heavy elements and surrounded by planetary debris is a unique tool for validating different models describing the formation and evolution of terrestrial-type planets (e.g., Harrison et al. 2018). Given the favorable planet-to-WD radius ratio, any terrestrial planet that might be detected will immediately become a prime target for atmospheric classification and bio-signature searches (e.g., Loeb & Maoz 2013). While searching for exoplanet transits around WDs, LAST is expected to find even more systems with transiting debris. Given its much higher cadence compared to ZTF, LAST will be able to put better constraints on the occurrence rate of such systems. Comparing the occurrence rate as a function of the WD cooling age would also help constraining the lifetime of such debris disks.

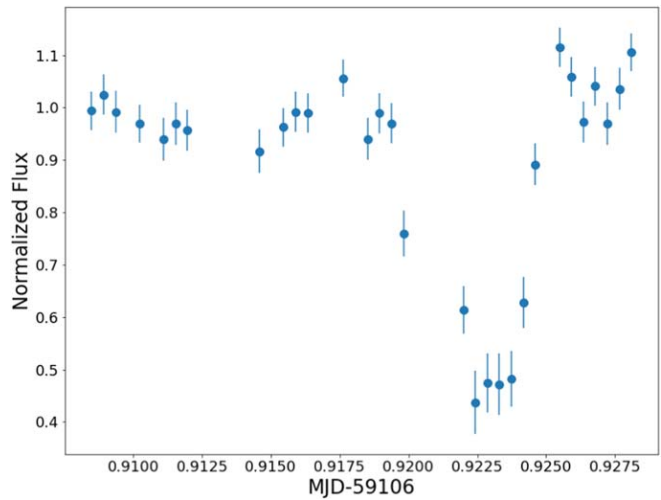

**Figure 9.** A transit of the WD1856+534 by an exoplanet candidate (Vanderburg et al. 2020) taken by a single LAST telescope in a test run on 2020 September 14.

LAST approach for WD exoplanet transit search is based on the conclusions derived by previous attempts to detect habitable-zone (HZ) planets orbiting WDs (e.g., Faedi et al. 2011; Fulton et al. 2014; van Sluijs & Van Eylen 2018): Covering a FoV as large as possible to increase the number of WDs monitored, and prioritizing cadence over photometric precision. Given the S/N measurements using a single LAST telescope (Figure 3), we expect to be able to detect Earth-sized transits with an impact factor $b = 0$ down to a magnitude 19.6 in $G_{bp}$ (with $G_{bp} - G_{rp} \approx 0.5$). For a significant number of WDs, we will be able to detect transits of smaller objects such as disintegrating planets and/or partial transits.

To understand what is the optimal cadence per night for a WD planet search with LAST, we run a Monte Carlo (MC) simulation in which we vary the total amount of time observed per night, and calculate the detection probability following a month of observations. We assume that the orbits are spread uniformly over a range of $6 \, \text{hr} \leqslant P_{\text{orb}} \leqslant 48 \, \text{hr}$, and require two observed transits to mark a detection. We then integrate over the orbital range and multiply by the number of fields observed per night given the accumulated exposure time. A maximal detection efficiency is achieved for 45 minutes exposures per field per night, which encouraged us to set the LAST high cadence mode to this value, see Figure 10. The total observing time required to monitor ≈50,000 WDs using such a strategy is

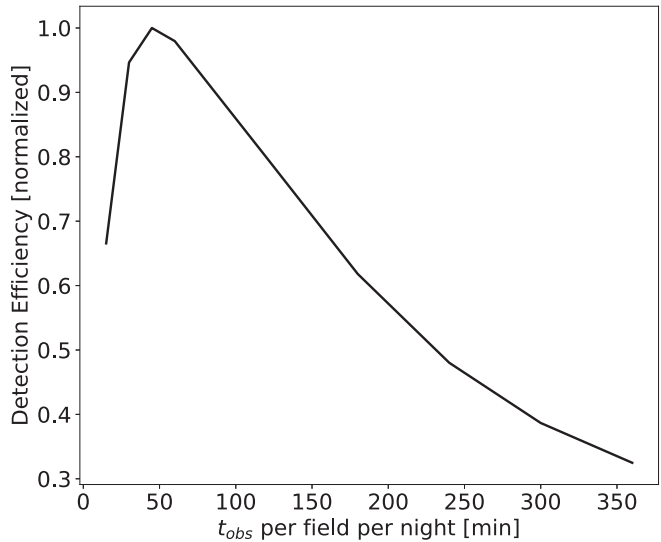

**Figure 10.** Optimal cadence for the WD transiting planet search with LAST: Normalized detection efficiency, taking the number of observed fields per night into account. See text for more details.

≈1800 hr. In case of an occurrence rate of 1% for Earth-sized terrestrial planets orbiting in the HZ of WDs (i.e., a factor of ∼25 below the occurrence level of terrestrial planets around MS stars, and significantly below the current upper limits of the occurrence rates of planets around WDs), we expect to detect about 5 terrestrial planets during the survey lifetime. Our effort will allow us to set an order-of-magnitude tighter constraints on the occurrence rate of HZ planets and minor objects around WDs.

### 5.2.3. Pulsations

As WDs cool, they pass through instability regimes where partial ionization of their dominant atmospheric constituents supports non-radial pulsations, resulting in measurable photometric variations at the system eigenfrequencies (Córsico & Althaus 2016; Córsico et al. 2016). For hydrogen-dominated (DAV) WDs the instability strip is at a temperature range of $10{,}800\,\mathrm{K} \lesssim T_\mathrm{eff} \lesssim 12{,}300\,\mathrm{K}$, while for helium-dominated (DBV) WDs it occurs at higher temperatures of $T_\mathrm{eff} \approx 25{,}000\,\mathrm{K}$ (e.g., Doyle et al. 2017). The pulsation amplitude is of the order of 10 mmag, with periods of 100–1000 s. Precision measurements of these modes, and mapping of the various eigenfrequencies versus other observables such as $T_\mathrm{eff}$ and log $g$ will allow us to study the interior structures of these WDs in detail (e.g., Metcalfe 2005; Giammichele et al. 2018). For several (∼5) coadds, we expect to detect pulsations down to 18 mag in case of large amplitudes ∼50 mmag. From the Gaia DR2 WD catalog photometric temperatures and model fits, we estimate ∼1000 pulsating WDs of both DAV and DBV spectral types to be monitored by LAST high cadence survey. Pulsating WDs in eclipsing systems provide even more valuable information, as their masses and radii can be constrained precisely through their eclipses. To date, only one pulsating WD in an eclipsing system is known (Hallakoun et al. 2016; Parsons et al. 2020; Romero et al. 2022). The LAST high cadence survey is expected to find more of these valuable systems. Assuming 10% of the WDs in the instability strip (Gianninas et al. 2015) are in double-WD binaries, and 1% of those are eclipsing, we expect to detect $\mathcal{O}(10)$ such systems with LAST.

### 5.3. Microlensing

Gravitational microlensing of stars by other objects in our Galaxy is a unique method to study several important populations of dark and hard to detect objects (see, e.g., review by Gaudi 2012). Among these are cold exoplanets (near and outside the snow line, e.g., Shvartzvald et al. 2016), exoplanets in different Galactic locations and environments, free-floating planets (e.g., Mróz et al. 2017), isolated stellar-mass BHs (Lam et al. 2022; Sahu et al. 2022), isolated low-mass brown dwarfs, and the so-called "brown-dwarf desert" (e.g., Ryu et al. 2017—a low mass brown dwarf discovered in a binary system). The access to these populations is also crucial for the study of Galactic structure and evolution.

Traditionally, microlensing surveys have concentrated their observing efforts toward the Galactic bulge (e.g., Udalski et al. 1992; Bond et al. 2001), where the microlensing event rate is highest. The Korean Microlensing Telescope Network (KMTNet; Kim et al. 2016) is currently the most efficient microlensing survey, with 24 hr coverage of 96 square degrees in the Galactic bulge and cadences ranging from 4 to 0.2 hr$^{-1}$, detecting ∼3000 microlensing events per year. Galactic-plane microlensing surveys, albeit with lower event rate, can provide important and complementary perspective to bulge surveys (Gould 2013). Moreover, the events are expected to have longer timescales ($t_\mathrm{E} \sim 60$ days), larger Einstein radii, and brighter magnitudes. Thus, facilitating the measurement of both the annual parallax and the astrometric microlensing signals, which together allow full determination of the physical properties of the system. For example, the only microlensing planet detected in the Galactic plane, TCP J05074264 +2447555 (Nucita et al. 2018; the so called "Kojima" event), was followed by both Spitzer and VLT-Gravity (Zang et al. 2020), as well as adaptive optics (AO) observations with Keck at high spectral resolution (Fukui et al. 2019), to fully determine the physical properties of the system.

Gould (2013) estimated the Galactic plane microlensing event rate, and found that a few hundred microlensing events per year are excepted to be detected in the inner ≈2000 deg$^2$ of the Galactic plane ($|l| \lesssim 90$, $|b| \lesssim 10$), and a few tens of events per year in the outer regions (over ≈12,000 deg$^2$). The former estimate is empirically confirmed by the derived event rate found by OGLE in their low-cadence Galactic plane survey fields (Mróz et al. 2020b). Additional low-cadence (few days) surveys search and detect Galactic-plane microlensing events, e.g., ZTF (Mróz et al. 2020a) and Gaia (Wyrzykowski et al. 2020; Rybicki et al. 2022). A low-cadence Galactic plane survey was also proposed to be conducted with the Vera C. Rubin Observatory (e.g., Gould 2013; Street et al. 2018; Sajadian & Poleski 2019), although its survey strategy has not been finalized yet.

LAST has a huge potential to lead Galactic-plane microlenisng surveys. For example, the high cadence survey strategy allows to cover the entire inner Galactic plane region ($\approx 2000\,\mathrm{deg}^2$), monitoring hundreds of events with an hourly cadence. This cadence is excellent for event detection, but moreover it is sufficient for anomaly detection and characterization due to the events' long timescales. The current LAST northern site is not optimal for carrying out such a survey. Field visibility imply that only 1/3 of the inner Galactic plane region is observable for at least 50 days (required to sufficiently cover a typical Galactic plane event), yielding an expected event rate of $\sim 50$ events $\mathrm{yr}^{-1}$. An hourly cadence from a single site will be sufficient to detect and characterize planetary anomalies due to massive planets, but these are expected for only $\sim 1\%$–$2\%$ of the events (Shvartzvald et al. 2016). A site in the southern hemisphere can make LAST an excellent tool for Galactic-plane microlensing survey, detecting $\sim 250$ events $\mathrm{yr}^{-1}$ and likely a few massive planets per year. A network of three southern LAST sites will allow 24 hr coverage, increasing the detection efficiency also to lower-mass planets.

All of the events detected in the LAST fields should also have Gaia data. For some of these events, Gaia is expected to detect the astrometric microlensing signal which enables to break degeneracies in the standard microlensing model, and derive their lens mass (e.g., Rybicki et al. 2018). However, the astrometric microlensing signal alone is insufficient, as it is necessary to have a good photometric coverage for these events to recover the mass of the lens. LAST has the potential to improve the light curves of microlensing events detected by Gaia, which are known to suffer from low cadence, insufficient for accurate characterization of the lens and source properties. Gaia is detecting microlensing events all over the sky and, for disk events, there is often limited survey data specifically around the asiatic gap—see Figure 4. While some of the events will get an additional follow-up (Hodgkin et al. 2021), not all will be found in real time and alerted. Moreover, even the baseline coverage might be useful, e.g., to assess the potential variability of the source, or identify a transient as a genuine microlensing event. It might also be interesting to check for lenses toward potential over-densities (and thus regions with higher microlensing rates), for example the Gould Belt, as suggested by Wyrzykowski et al. (2023). We take the opportunity to emphasize again that one of the key challenges is to obtain the manpower and funds, and implement the tools required for using LAST as a microlensing experiment, as well as for other science cases discussed throughput this publication (e.g., ATLAS forced photometry server; Shingles et al. 2021), which we further discuss in Ofek et al. (2023) (see also https://github.com/EranOfek/AstroPack).

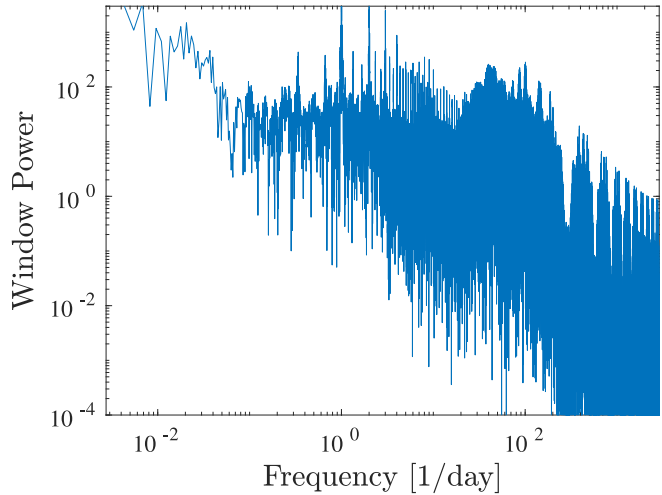

**Figure 11.** The window function of simulated scheduler observations, for a single field observed at high cadence ($\approx 8$ visits per night), in the frequency range of $1/180\,\mathrm{day}^{-1}$ to $86400/30\,\mathrm{day}^{-1}$.

### 5.4. Short Period Variables

With its $20 \times 20$ s exposures strategy, LAST will provide a census of short-period variables over $3\pi$ steradians to a limiting magnitude of about $\sim 19$, and will allow us to probe a new phase space of variable stars which until now was typically explored over a much smaller FoV. Among the known phenomena with periods in the range of tens of seconds to tens of minutes, are rotating WDs in intermediate polars (e.g., Patterson et al. 2020), stellar (non-radial) pulsations (e.g., Gautschy & Saio 1995, 1996), AM CVn stars (e.g., Ramsay et al. 2018), and NS/BH-WD binary systems (e.g., Toonen et al. 2018; Wiktorowicz et al. 2019). Among the known minutes-timescale transients, we can find flares from UV Cet stars (e.g., Haisch et al. 1991), and accreting BHs (e.g., Kasliwal et al. 2008; Hynes et al. 2019). These transients require continuous observations of the same field (e.g., EvryScope, Law et al. 2015; NGTS, Wheatley et al. 2018).

In order to examine the capabilities of LAST in terms of short-period variables, we performed simplistic (i.e., not taking into account weather) scheduling simulations of LAST over one year. In total, in such a simulation about 8900 individual images are obtained in about 440 visits. Figure 11 presents the window function of such observations, for a single field, in the frequency range of $1/180\,\mathrm{day}^{-1}$ to $86400/30\,\mathrm{day}^{-1}$.

A source of false positives of such events identified recently by Corbett et al. (2020) and Nir et al. (2021a, 2021b) showed that geostationary satellites produce a high-rate foreground of glints with sub-second duration that have a stellar PSF. Nir et al. (2021a) estimated the rate of these flares to be above 30 deg$^{-2}$ day$^{-1}$ for events brighter than magnitude 11 (corresponding to magnitude ∼16, when diluted by LAST 20 s exposure time). LAST, when observing with a strategy of 20 × 20 s exposures per visit, will be able to distinguish between satellites glints and transients with ≳20 s duration.

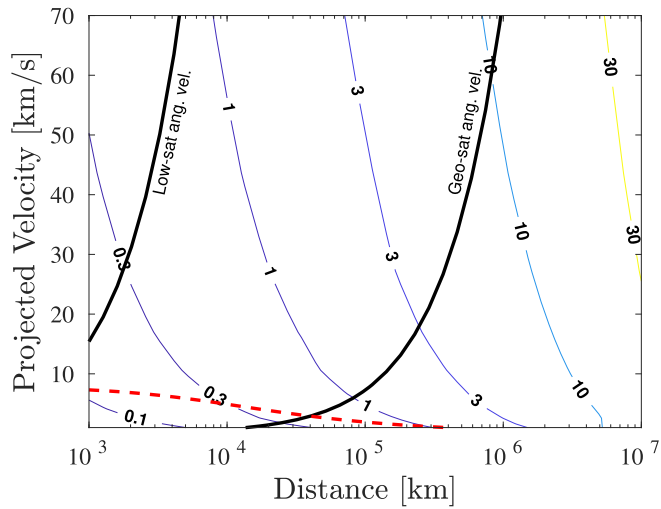

**Figure 12.** Contours show the minimum-radius, in meters, of NEAs detectable using streak detection, as a function of their distance and projected velocity. The calculation assumes 10% albedo, and that the streak is detected in a single image from a single telescope and the streak length is limited to 1024 pix. The bold-black lines present equal angular velocity lines for the angular velocity of geostationary satellites (15″ s$^{-1}$) and low-orbit satellites (3200″ s$^{-1}$). The red dashed line shows the Keplerian velocity as a function of distance (from a topocentric observer), of an object in a circular orbit around the Earth. Since factors like phase function are not included in this plot, it should be regarded as a rough approximation.

## 6. Solar System

### 6.1. Near-Earth Asteroids

LAST will be a valuable addition to the multitude of sky surveys monitoring Near-Earth Asteroids (NEAs) due to its large grasp and geographic location, see Figure 4. NEAs impact the Earth and can pose local danger approximately once a century (Harris et al. 2015). In addition, sub-km size NEAs passing at fractions of astronomical unit from the Earth, are a sole source of information about small-sized asteroids, in terms of numbers, size distribution, shape, spin state and correlations between these parameters. Photometric measurements of these bodies are the most common tool to drive these properties (e.g., Polishook et al. 2012; Thirouin et al. 2016). LAST large FoV and high cadence will allow us to determine these parameters for bright NEAs. In addition, using the streak detection method (Nir et al. 2018), LAST large FoV will allow us to detect more than 150 NEAs with Vmag values brighter than the limiting magnitude (<19.4). Figure 12 shows the minimum-radius, in meters, of NEAs detectable using this method, as a function of their distance and projected velocity. Collecting these NEAs' astrometry will benefit constructing their orbits by the Minor Planet Center, while their photometric measurements, even at a low number of nine visits stretching over ∼5 hr, will yield a crude estimate of their light curves, rotational periods and elongation ratio.

About 20 of the observed NEAs will have diameters smaller than ∼150 m (absolute magnitude 21 in H band). Such small NEAs were found to rotate orders of magnitude faster than larger asteroids, reaching rotation period records of ∼10 s (e.g., Thirouin et al. 2016), and most probably are being pushed to such fast rotation by thermal torques caused by the YORP effect (Bottke et al. 2002). LAST high cadence of ∼20 × 20 s exposures, will be an effective tool to identify NEAs with sub-minute rotation periods, measure their numbers, and confirm or refute their correlation with the YORP effect.

Given the large number of debris and satellites in Earth orbit, one challenge is to distinguish between a satellite and an NEA. The simplest solution is to restrict the search to objects with angular velocity below ∼15″ s$^{-1}$ (i.e., the angular velocity of geostationary satellites). A method that will allow us to distinguish between a satellite and an NEA is observing a deviation from uniform velocity. For a NEA at distance of 40,000 km, moving tangentially at 5 km s$^{-1}$, an half-an-hour arc is needed in order to detect deviations from constant angular speed.

### 6.2. Main Belt Asteroids

Time-series photometry of asteroids is used to derive their sizes, shape models, rotation periods, spin-pole vector, and the existence of satellites (e.g., Durech et al. 2015; Li et al. 2015; Margot et al. 2015). These parameters are fundamental to our understanding of asteroids' internal structure, tensile strength and evolution (e.g., Scheeres et al. 2015).

*Numbers and lightcurves:* At any given moment there are about $2 \times 10^4$ observable asteroids brighter than magnitude 19. Using LAST high cadence and wide FoV, about 2800 asteroids can be observed per night. With this approach, six-hour long light curves (with 7 gaps within) will be collected per asteroid, which is the mean rotation period of ∼10 km sized asteroids. Additional light curves per asteroid will be collected every ≈10 days, which allows deriving the spin rate of slow rotators. With the low-cadence approach, 3 to 4 times more asteroids will be observed, shorter daily light curves will be collected (∼12 minutes per night), but each asteroid will be observed every three nights (resulting in ∼200 points per month, taking the Moon into account). This approach will result in less rotation-period values, but similar numbers of light curve amplitudes. Annually, LAST can deliver light curves for ∼$10^5$ asteroids.

*Shape models:* A large number of photometric points at a low cadence can be used to match shape models and spin vectors through light curve inversion techniques (e.g., Durech et al. 2015), especially when the asteroids are observed at varied geometric combinations. Shape models enable volume calculations (e.g., Carry 2012); different patterns of asteroid shapes are the result of asteroid evolution (e.g., Walsh et al. 2008); the shape-model solution is also constrained by the spin vector, which is in turn affected by the YORP effect (e.g., Vokrouhlický et al. 2003) or by mutual events with existing satellite (e.g., Pravec et al. 2006).

*Binary asteroids:* Satellites of asteroids can be detected using photometry in the cases of eclipses and occultation events between the two components of the system. The light attenuation can be short in case of partial mutual events, or reach ∼10% of the orbital period (e.g., Polishook et al. 2011). About one out of six asteroids have satellites (Margot et al. 2015), with orbital periods of a few tens of hours. This means that there is a larger chance to detect satellites on the high-cadence survey, and specifically in the cases of asteroids observed during multiple nights, once per ∼10 days. Therefore, we estimate that a few hundred binary asteroids could be identified annually by LAST. Another method for detecting binary asteroids is via astrometry (Segev et al. 2023).

*Structure:* Asteroids' "rubble pile" structure does not allow them to rotate at a higher rate than 11 revolutions per day (e.g., Scheeres et al. 2015). Some exceptions provide opportunities to learn about additional forces acting on asteroids constituents, such as cohesion and friction (e.g., Polishook et al. 2016). Small-sized asteroids ($\lesssim$100 m) that can cross the rubble-pile spin barrier are most probably monolithic in nature (e.g., Scheeres et al. 2015), and are most probably the components constructing the larger rubble-pile asteroids. LAST sampling rate of $20 \times 20$ images is sensitive to both groups, and increases our knowledge of asteroids' structure and spin evolution.

*Collisions:* Collisions between asteroids are one of the key mechanisms that control their dynamical and structural evolution (e.g., Davis et al. 2002; Holsapple 2022, and reference therein). Collisions result in ejecta activity that last on timescales of hours to months, and can be spotted by an increase in brightness or by a resolved light source (e.g., P/2010 A2—Jewitt et al. 2010, 596 Scheila—Jewitt et al. 2010). Artificial impacts, such as the DART impact mission on the asteroid Dimorphos (Rivkin et al. 2021), observed by the LAST telescopes both in real time and in the following months, suggest that such events seen at a distance of ∼2 au will be detected as 1 hr long transients, followed by a fainter a-few-week-duration transient. Detecting such impact events will provide us the information to study the breakup physics, collision rate, asteroid size distribution, and zodiacal dust production.

### *6.3. Kuiper Belt Objects*

The distribution of objects in the outer solar system provides important clues to its formation and evolution. For example, the existence of Sedna-like objects (Brown et al. 2004; Trujillo & Sheppard 2014) is not well understood, and cannot be explained as the result of scattering by bodies in the known parts of the solar system (e.g., Gladman et al. 2002; Gomes 2003). Among the explanations are perturbations by close encounters with passing stars (e.g., Brown et al. 2004), and the formation of the Solar system in a star-cluster environment (e.g., Fernández & Brunini 2000). Another intriguing explanation, supported by the analysis of orbits of known Kuiper-belt objects, hints on the existence of a yet unknown massive planet ($\gtrsim$10 $M_\oplus$; Batygin & Brown 2016).

To date, the deepest wide-field searches for distant KBOs reach a limiting magnitude of 21.3 over a fraction of the sky (Schwamb et al. 2009). Deeper surveys, to limiting magnitude of 23, were performed, mostly in the southern hemisphere, but for a FoV of 5000 deg$^2$ (e.g., Trujillo & Sheppard 2014; Sheppard et al. 2016; Bernardinelli et al. 2020).

LAST can reach a limiting magnitude of ∼23 (equivalent to KBOs with a radius of ∼100 km) by stacking images collected by a single telescope during ≈7 hr from a single FoV. The minimal angular velocity of KBOs at quadrature (Earth-Sun-field angle of ∼90°) is ≈0$''$002 minute$^{-1}$. Long observing times, that will keep a KBO with a round PSF under the seeing limitations of the observatory and will allow us to reach the aforementioned magnitude, is possible during local winter. Observing the same FoV again in the following night, will allow to detect its motion. Using the wide observing mode (of ∼350 deg$^2$), LAST can cover the Kuiper-belt arc in the sky in seven to ten nights. With ∼100 known KBOs brighter than 23 mag at any given time around the quadrature, such a campaign is guaranteed to yield significant results.

### *6.4. Occultation by Oort Cloud Objects*

The Oort Cloud (Oort 1950) is a primordial relic of the solar system formation. So far there are no direct observations of objects in the Oort Cloud, and their numbers, distances, size distribution, and spatial distribution are highly uncertain. The content of the Oort Cloud may hold some clues regarding the solar system formation and evolution (e.g., Heisler & Tremaine 1986; Fouchard et al. 2018). Bailey (1976) suggested that it will be possible to detect small Trans-Neptunian objects by occultations. However, due to their tiny cross sections and short durations, these occultations are hard to find. So far only three occultations by Kuiper-Belt objects have been reported (Schlichting et al. 2009, 2012; Arimatsu et al. 2019).

Oort cloud occultations are harder to detect, because the Fresnel angular scale is smaller than the angular radius of Sun-like stars with an apparent magnitude of 15.5. Given the sub-second duration of these occultations, this requires large telescopes (e.g., $\gtrsim 1$ m). However, E. O. Ofek & G. Nir 2023, in preparation (see also Brown & Webster 1997) suggested that observing near the quadrature points with a few-second integration time may be a more effective method to detect Oort cloud objects compared to sub-second integrations at opposition. The reason for this is that while the duration of the occultations near quadrature increases, and hence the rate of the occultations decreases (by about an order of magnitude), the increased duration means that we can use longer integrations that in turn increase the number of monitored stars, whose angular size is smaller than the Fresnel scale, by about two orders of magnitudes. LAST, with its wide FoV, will allow us to monitor many stellar objects with a Fresnel scale smaller than that of Oort cloud objects, compensating for the lower occultation rate—in an attempt to detect Oort cloud objects for the first time. This strategy requires ∼1 s integration time—in accordance with LAST capability of taking 0.8 s exposures with practically no dead time.

## 7. Summary

LAST is a unique wide-FoV survey with a modular design allowing it to adopt various survey strategies, many of them simultaneously. The scientific capabilities of such a facility are broad and encompass all aspects in modern astrophysics, from the study of GW sources to the study of planetary systems and stellar remnants in the Milky Way, and the search for asteroids and NEAs in our own backyard.

LAST unprecedented grasp, comparable only to large-scale facilities which require capital investment an order of magnitude larger, is achieved through the use of COTS components that became available at relatively low costs in the last couple of years. The facility can therefore be readily duplicated to multiple sites around the globe. A significant advantage of LAST is that its grasp is driven by the FoV rather than depth, and allows follow-up of potential discoveries with small- and medium-sized telescopes.

As mentioned in the introduction, in order to increase the utility of the survey, the data will be publicly released every ∼12 months. The authors encourage potential collaborators that would like to take advantage of the experience gained in building the facility, to contact them in an effort to build a network of such observatories that will boost the capabilities of the individual facilities. As the bottleneck of modern sky surveys is often in the availability of manpower and funds, collaboration with the scientific community will determine whether the scientific aspirations detailed in this publication will become a reality.

## Acknowledgments

S.B.A. is grateful for support from the Azrieli Foundation, André Deloro Institute for Advanced Research in Space and Optics, Peter and Patricia Gruber Award, Willner Family Leadership Institute for the Weizmann Institute of Science, Aryeh and Ido Dissentshik Career Development Chair, Israel Science Foundation, Israel Ministry of Science, and Minerva Stiftung.

E.O.O. is grateful for support by grants from the Willner Family Leadership Institute, Madame Olga Klein—Astrachan, André Deloro Institute, Schwartz/Reisman Collaborative Science Program, Paul and Tina Gardner, The Norman E Alexander Family M Foundation ULTRASAT Data Center Fund, Jonathan Beare, Israel Science Foundation, Isreal Ministry of Science, Minerva, BSF, BSF-transformative, and the Rosa and Emilio Segre Research Award.

I.A. is a CIFAR Azrieli Global Scholar in the Gravity and the Extreme Universe Program and acknowledges support from that program, from the European Research Council (ERC) under the European Union's Horizon 2020 research and innovation program (grant agreement No. 852097), from the Israel Science Foundation (grant No. 2752/19), from the United States—Israel Binational Science Foundation (BSF), and from the Israeli Council for Higher Education Alon Fellowship.

A.F. and V.F.R. acknowledge support from the German Science Foundation DFG, via the Collaborative Research Center SFB1491: Cosmic Interacting Matters—from Source to Signal.

N.L.S. is funded by the German Science Foundation (DFG) via the Walter Benjamin program—461903330.

## ORCID iDs

S. Ben-Ami https://orcid.org/0000-0001-6760-3074
N. Hallakoun https://orcid.org/0000-0002-0430-7793
E. Segre https://orcid.org/0000-0002-0913-3083